# A STATISTICAL REAL-TIME PREDICTION MODEL FOR RECOMMENDER SYSTEM

Team Name: IITDU_Espresso
Email: minhaskamal024@gmail.com

Md Rifat Arefin
Institute of Information Technology
University of Dhaka
rifat.arefin515@gmail.com

Minhas Kamal
Institute of Information Technology
University of Dhaka
minhaskamal024@gmail.com

Kishan Kumar Ganguly
Institute of Information Technology
University of Dhaka
kkganguly.iit.du@gmail.com

Tarek Salah Uddin Mahmud
Institute of Information Technology
University of Dhaka
tsmrafee@gmail.com

## ABSTRACT
Recommender system has become an inseparable part of online shopping and its usability is increasing with the advancement of these e-commerce sites. An effective and efficient recommender system benefits both the seller and the buyer significantly. We considered user activities and product information for the filtering process in our proposed recommender system. Our model has achieved inspiring result (approximately 58% true-positive and 13% false-positive) for the data set provided by RecSys Challenge 2015. This paper aims to describe a statistical model that will help to predict the buying behavior of a user in real-time during a session.

## Categories and Subject Descriptors
H.2.8 [**Database Management**]: Database Applications – *data mining, statistical databases*; H.3.3 [**Information Storage and Retrieval**]: Information Search and Retrieval– *relevance feedback.*

## 1. INTRODUCTION
With the advancement of e-commerce, online shopping is becoming increasingly popular now-a-days. These e-commerce sites contain millions of products of thousands of brands. It is really challenging for customers to seek and select necessary products among so many of those options. Here comes the recommender system. It receives data about a consumer's behavior, predicts his/her need and attempts to help him/her in choosing a product. [1]

The main problem in a recommender system is to balance between scalability and performance. [2, 3] It is also hard in maintaining good result for both new users with very limited information and old users having lots of buys and ratings [3]. In the problem of RecSys Challenge 2015, we have a sequence of click events performed during a typical browsing session in an e-commerce website. By processing this, we need to predict that- if there will be any buying event in a session and which items are going be bought.

Some very well-known and common approaches toward solving this problem are- Collaborative Filtering, Content-Based Filtering [6], Rule-based Modeling, Cluster Model, Search-Based Method, Item-to-item Correlation [3, 4], and so on [8, 9].

Most popular recommender systems implement Collaborative Filtering. [2, 3] However this technique has a fundamental trade-off problem between scalability and accuracy. With the increase of popularity of ecommerce sites, it has becomes even harder in finding relations among billions of users and representing the result at real time.

Rule-based modeling may be regarded as a good answer for previous problem. But its success rate depends mostly on the set of rules that it work with. So, after a very careful data analysis an optimal situation may be reached. However with the rapid evolution of technology and change in customers' requirements, this state does not last long. And analyzing data every time is both costly and time consuming.

Cluster Model is an excellent solution [2]. It distributes all customers to many small segments depending on their similarities of preferences. When a new user arises, the system attempts to assign the user to a group. And then it recommends him/her depending on that group. But assigning customer to a suitable group is a big challenge. Failure of proper distribution may result in annoyed customers.

Item-to-item correlation recommender scheme is a wonderful resolution [3, 4]. Rather processing long consumer history, it works usually with current interest. Thus user gets recommendation over recent hot products without having any past record. But this may not be suitable for all users.

To overcome these problems we have used two-step filtering which can deliver result in real-time. We extracted two types of information from the training data- consumer side information (i.e. number of clicks, duration) and product side information (i.e. product popularity). The test data was primarily filtered with consumer side information and then with product side information. We could not implement the full extent of our model, yet it scored 42287.9 in the leader board.

## 2. METHODOLOGY

We propose a two-step solution to solve the RecSys Challenge 2015. In step-1, we calculate the likelihood ratio of an item being bought or not-bought based on a set of selected features and compare it against a threshold. The output of step-1 is then fed to the step-2 for the refinement purpose. In step-2, we use item popularity (defined below) and compare it against a threshold to get the final output.

**Feature Selection**

Analyzing the training data files provided by RecSys, we take following features into account- hour of day, date of month, day of week, month of year, number of clicks on an item in a session, session duration, etc. We will discuss how these features effect on the buying behavior of a user.

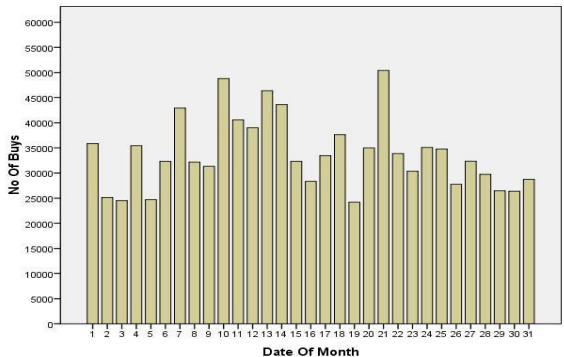

Figure 1. Date-Buy Relationship

From Figure 1, we get that- no. of buys varies on the difference of dates of a month. For example: people tend to buy things more on date 10, 13, 14, and 21 than other date of a month. After date 21 buying events reduce.

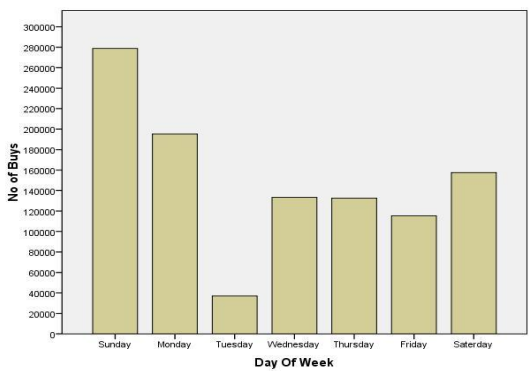

Figure 2. Day of Week vs. Number of Buys

From Figure 2, we understand that- customers have maximum number of buy events on Sunday and minimum on Thursday.

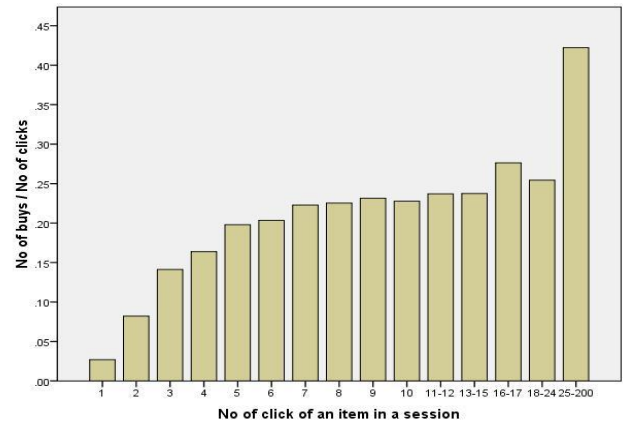

Figure 3. Clicks vs. Probability of Buy

In Figure 3, we see that- as the no. of clicks on an item in a session increases, the probability of buy increases too.

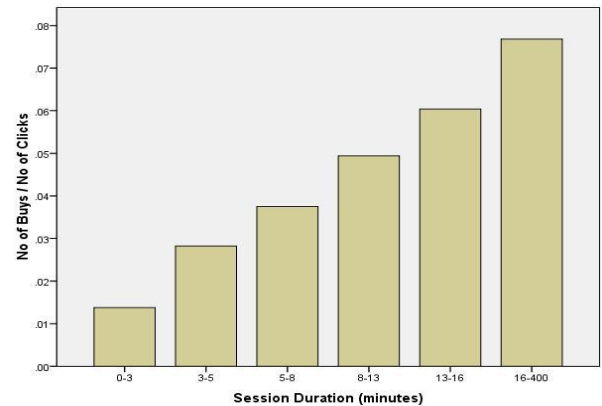

Figure 4. Session Duration vs. Probability of Buy

Session duration (in minutes) is the time duration between $1^{st}$ and last click in a session. From Figure 4, we see that- with the increase of session duration the probability of buying an item in that session also increases.

After analyzing the data we see that there is a reasonable separation between values of features of buy and non-buy items. This fact can be used to build a classifier which will be used in step-1 filtering.

**Item popularity**

To filter out the output of step-1 we use popularity of each item which is defined using the following equation:

$$\text{Popularity (p)} = \frac{b}{c}$$

Here,
b = No. of buys of an item
c = No. of clicks of an item

In step-2 filtering, we categorize popularity in three sections- low, medium and high. We multiply the no. of clicks of each item with its popularity before comparing against a threshold. So, if a customer clicks more on an item then the chance of buying that will increase.

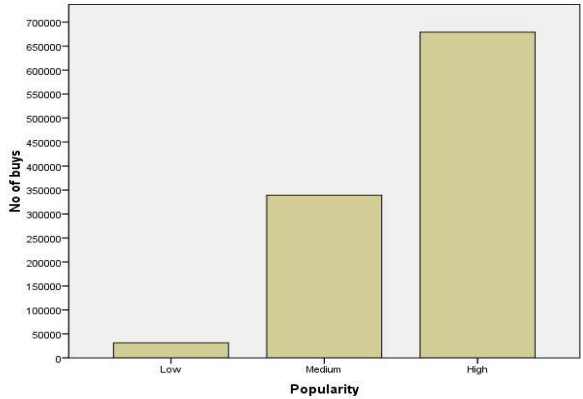

Figure 5. Popularity vs. Number of Buys

**Filtering process**

Let, prediction outcome be- 'Buy' or 'Non-Buy'; and features of an item are $x_1, x_2, x_3 \ldots x_n$. Given items as buy or non-buy, the probability (conditional probability) that a given item belongs to buy and non-buy is defined as:

$$P(x_1, x_2, x_3 \ldots x_n | Buy) = \frac{B(x_1, x_2, x_3 \ldots x_n)}{T_b}$$

$$P(x_1, x_2, x_3 \ldots x_n | NonBuy) = \frac{N(x_1, x_2, x_3 \ldots x_n)}{T_n}$$

Here,

$B(x_1, x_2, x_3 \ldots x_n)$ = No. of buy items when item features are $x_1, x_2, x_3 \ldots x_n$.

$T_b$ = Total buy items

$N(x_1, x_2, x_3 \ldots x_n)$ = No. of non-buy items when item features are $x_1, x_2, x_3 \ldots x_n$.

$T_n$ = Total non-buy items

Given the conditional probabilities of buy and non-buy items, a classifier can be built using Bayes maximum likelihood approach and it will be used for step-1 filtering. [5] Using this classifier items from test data will be filtered and selected for step-2 filtering if:

$$\frac{P(x_1, x_2, x_3 \ldots x_n | Buy)}{P(x_1, x_2, x_3 \ldots x_n | NonBuy)} > T_1$$

Here, $T_1$ is a predefined Threshold.

Algorithm-1 describes the step-2 filtering process using the popularity of items.

| Algorithm-1: Popularity Based Filtering Algorithm |
|---|
| **Input:** Items with their no. of clicks in its corresponding session |
| **Output:** Selected items as buy |
| **Begin** |
|     **For** each items selected from step-1 |
|         **If** item popularity exists |
|             p = popularity of the item, n = no. of clicks of the item in a session |
|             **If** (p * n > $T_2$) |
|                 Then select the item as buy |
|         **Else** |
|             Select the item as buy |
|     **End For** |
| **End** |

Here, $T_2$ is a predefined threshold.

**Why two step filtering?**

In our problem, we need to predict whether there will be a buy in a click session or not; and what item will be bought in that session. According to the problem, we divided our solution into two steps. In step-1 filtering-the main focus is on sessions with buying events. Items of the same category are selected in step-1 filtering. Customers tend to click different items of same category in a session. However the tendency of buying popular items is more. So we used step-2 filtering, where more popular items will be finally selected as bought. Our method produces a real time result because; we just compare the likelihood ratio and popularity against two thresholds and provide the result. This model can easily adapt with data increment, because computational complexity of training is also low.

## 3. RESULTS

We have performed several prediction experiments to see the runtime performance and scalability of our proposed method. There are around 9.2 million sessions in the click data. We have divided this data into train and test data, and have performed 5-fold cross validation. Its outcome is demonstrated is in Table-1.

**Table 1. Experiment Result**

| Test case | Score | False-Positive (%) | True-Positive (%) |
|---|---|---|---|
| T-1 | 33062.4 | 13.96 | 58.72 |
| T-2 | 32973.7 | 13.93 | 57.94 |
| T-3 | 32477.9 | 14.01 | 57.06 |
| T-4 | 33694.8 | 13.56 | 57.43 |
| T-5 | 32627.1 | 13.78 | 56.97 |

To match the challenge of Recsys, we also have taken 2.3 million sessions randomly from the click data (same amount of test data supplied by Recsys). Rest of the data (around 7.7 million sessions) is used as training data. It is noteworthy to mention that, in our test data the no. of buy also remain around 5%. Using the same scoring method given by Recsys, we scored 41406.7, while in the competition leader board we got 42287.9.

## 4. CONCLUSION

As recommender system provide a fully personalized marketing experience, its necessity is growing swiftly with the rapid growth of e-commerce. In this problem, we needed to predict buying behavior of customers. We tried to provide a statistical approach comprising of two-step filtering.

Our method offers a simple yet effective and practical solution to this. It is both efficient and scalable. It also produces reliable result even with scarcity of information. Although it has a small false positive percentage but that is high in number. By tuning the threshold this result can be enhanced immensely. Its performance can be further improved by including some more features (such as- item category, season wise item popularity, etc.) in the process.